\documentclass[journal, twocolumn]{IEEEtran}
\usepackage[T1]{fontenc}
\usepackage{color} 
\usepackage{amssymb}
\usepackage{mathtools}
\usepackage{algorithmic}
\usepackage{array}
\usepackage{mdwmath}
\usepackage{mdwtab}
\usepackage{bm}
\usepackage{mathrsfs}
\usepackage{epsfig}
\usepackage{subfigure}
\usepackage{algorithmic}
\usepackage{amsmath}
\usepackage{mathrsfs}
\usepackage{cite}
\usepackage{url} 
\graphicspath{{./fig/}}

\newcommand{\qW}{{\bf W}}
\newcommand{\qI}{{\bf I}}
\newcommand{\qw}{{\bf w}}
\newcommand{\qp}{{\bf p}}

\newcommand{\qq}{{\bf q}}
\newcommand{\qh}{{\bf h}}

\begin{document}

\title{Embedding Model Based Fast Meta Learning   for Downlink Beamforming Adaptation}
\author{
Juping Zhang, Yi~Yuan,
       ~Gan~Zheng,~\IEEEmembership{Fellow,~IEEE,}
       Ioannis Krikidis,~\IEEEmembership{Fellow,~IEEE,}
       and Kai-Kit Wong,~\IEEEmembership{Fellow,~IEEE}
\thanks{J. Zhang and G. Zheng are with the Wolfson School of Mechanical, Electrical and Manufacturing Engineering, Loughborough University, Loughborough, LE11 3TU, UK (E-mail: \{j.zhang3, g.zheng\}@lboro.ac.uk).}
\thanks{Y. Yuan was with   the Wolfson School of Mechanical, Electrical and Manufacturing Engineering, Loughborough University, Loughborough, LE11 3TU, UK  and is now with the 5G Innovation Center, Institute of Communication Systems, University of Surrey, Guildford GU2 7XH, U.K (E-mail: yy0007@surrey.ac.uk).}
  \thanks{I. Krikidis is with the    Department of Electrical and Computer Engineering,  University of Cyprus, 1678 Nicosia, Cyprus (E-mail: krikidis@ucy.ac.cy).}
 \thanks{K.-K. Wong is with the Department of Electronic and Electrical Engineering, University College London, London, WC1E 6BT, UK (Email: kai-kit.wong@ucl.ac.uk).}
 }
\maketitle
\vspace{-2cm}
\begin{abstract}
This paper studies the fast adaptive beamforming for the multiuser multiple-input single-output downlink. Existing deep learning-based   approaches assume that training and testing channels follow the same distribution which causes task  mismatch,   when the testing  environment changes. Although   meta learning can deal with  the task mismatch, it relies on labelled data and incurs  high complexity in the pre-training and fine tuning stages. We propose a simple yet effective adaptive framework to  solve the mismatch issue, which trains an embedding model as a transferable feature extractor, followed by fitting the support vector regression. Compared to the existing meta learning algorithm, our method does not necessarily need labelled data in the pre-training and does not need fine-tuning of  the pre-trained model in the adaptation.  The effectiveness of the proposed  method is verified through two well-known applications, i.e., the signal to interference plus noise ratio balancing problem and the  sum rate maximization problem. Furthermore,  we extend our proposed  method to online scenarios  in non-stationary environments.  Simulation results demonstrate the advantages of the proposed algorithm in terms of both  performance and complexity.  The proposed   framework   can also be applied to general radio resource management problems.
\end{abstract}

\section{Introduction}
Beamforming is one of the most promising multi-antenna techniques that can realize the antenna diversity gain and mitigate multiuser interference simultaneously. Optimizing beamforming weights is critical to fully reap its benefits and has been extensively studied in the literature  for various objectives  such as power minimization \cite{rashid1998transmit,bengtsson1999optimal},  signal-to-interference-plus-noise ratio (SINR) balancing \cite{schubert2004solution}, and sum rate maximization \cite{shi2011iteratively}. However, most beamforming solutions are highly complex to implement and cannot meet the critical latency requirement in the fifth generation (5G) and beyond systems    because they are iterative in nature and provide slow convergence.

Recently the deep learning technique has been proposed to address the complexity of beamforming design  using the `learning to optimize' framework \cite{sun2017learning}. It is based on the intuitive idea that the mapping from    channel state to beamforming can be learned by training a   neural network model in an offline manner, and then the beamforming solution can be directly predicted using the trained model in real time.  This method  shifts the complexity of real-time beamforming optimization to offline training   and its potential has been demonstrated in solving a series of beamforming design problems  \cite{alkhateeb2018deep,shi2018learning,huang2018unsupervised,xia2019deep,huang2019fast, dl_sumrate}. A major drawback of the existing deep learning-based beamforming solutions is that they are restricted to   static wireless environments, in which  the training and testing channels follow the same distribution; this assumption is  practically unrealistic. In practical wireless networks, the channel distribution may change due to   high mobility (e.g., in vehicular networks) or unexpected perturbations of   complex environments. Consequently, a well-trained model based on the training data could cause unacceptable performance degradation in the testing environment.  To tackle this challenge, an obvious solution is to re-train the model by using newly collected data from the new   environment. This is impractical because the changing network does not allow enough time to collect enough  new data and then train a new model before violating the latency constraint. Therefore it is a pressing research challenge in multi-antenna communications to achieve fast adaptation of beamforming solutions.

Transfer learning and meta learning have been recognized as two emerging techniques to design adaptive beamforming. Transfer learning \cite{pan2009survey} has been applied to improve performance of various resource allocation problems in wireless communications due to its strong ability of transferring   useful prior knowledge from the old scenario to a new one  \cite{zappone2019intelligence,shen2018transfer}.  In machine learning practice, fine-tuning  is a widely used transfer learning technique which re-trains part of a pre-trained model on a new but related task with new data.  Different from transfer learning, meta learning is a `learning to learn' strategy  which aims to train a model with a   better generalization ability \cite{thrun1998learning}.  It mainly includes metric-based  and optimization-based meta learning. The
metric-based meta learning aims to learn to  embed input data into a task-specific metric space in which learning of new tasks is efficient. Examples of embedding include  Siamese networks, matching networks, relation networks and prototypical networks \cite{tian2020rethinking}. The optimization-based meta learning aims to learn the gradient-based neural network optimization such as hyper-parameters so that the neural network can effectively learn new tasks. One of the most popular optimization-based meta learning algorithms is    the model-agnostic meta-learning (MAML) algorithm proposed in \cite{finn2017model} that is widely applied to general deep neural networks. The MAML algorithm aims to learn a parameter initialization of the deep neural network model, such that a small number of gradient updates of the parameter by using limited testing data from a new task will produce large reduction in the loss function of that task.   The MAML algorithm   has been successfully  used to deal with the end-to-end   decoding problem over fading channels \cite{park2019learning} \cite{park2019meta}  with few pilots and to learn the downlink channel state information (CSI) from the uplink CSI in frequency division duplexing (FDD) systems \cite{yang2019deep}. Our previous work in \cite{MAML_BF} has also demonstrated that MAML-based adaptive beamforming can rapidly adapt to the new environment and is superior to the transfer learning method.

Although the MAML algorithm has shown a good ability on solving the mismatch issues of beamforming in designing dynamic wireless networks \cite{MAML_BF}, it may cause outdated beamforming prediction due to its high complexity in the training and adaptation processes. Hence, it is important to design a new method, which can reduce the complexity and maintain the performance of adaptive beamforming.
 The intuition of our proposed approach  is based on the findings in \cite{tian2020rethinking} and \cite{raghu2019rapid} that the dominant factor of the effectiveness of the MAML algorithm  is the feature reuse   rather than the sophisticated design of the meta learning algorithm. Motivated by the results in  \cite{tian2020rethinking} and \cite{raghu2019rapid}, the novelty of this paper is to propose  a new simple and effective embedding-based meta learning algorithm to extract the main features and improve the MAML algorithm for adaptive beamforming by reducing the complexity while guaranteeing the efficiency and performance.  Our main contributions are summarized as follows:
\begin{itemize}
\item We propose a simple yet effective   learning framework for adaptive beamforming design, which first trains an embedding model by using existing data in the pre-training stage, and then fits a new model using support vector regression (SVR) with limited new channels and labelled solutions in the adaptation stage. Our proposed method is  based on the idea of embedding in the metric-based meta learning, so it is different from the transfer learning method. In addition,    different from the existing transfer learning and meta learning methods, our proposed design framework is applicable to both supervised and unsupervised learning in the pre-training stage, and it trains a simple regression model instead of fine-tuning the pre-trained model in the adaptation stage.
\item We apply the proposed framework to design specific adaptive beamforming algorithms for the  SINR  balancing problem  and the sum rate maximization problem, which use  supervised learning and  unsupervised learning, respectively.
\item  To further investigate the effectiveness of the proposed fast learning framework on beamforming design in non-stationary scenarios, we extend our framework to the online application to solve the beamforming prediction problem in   real-time communications systems.
\item Extensive simulations are carried out to assess the adaptation capability of the proposed algorithms in realistic communications scenarios. The results verify that the proposed adaptive beamforming algorithms improve the adaptation performance, the stability  and the computational efficiency in both pre-training and adaptation stages compared to transfer learning and MAML.
     \end{itemize}

The remainder of this paper is organized as follows. Section \ref{system_model} introduces the system model, problem formulation and the existing MAML algorithm. In Section \ref{fastalgorithm}, the proposed fast meta learning framework and the algorithms for the two applications are presented. Section \ref{online_adaptation} develops the fast online meta learning based adaptation algorithm. Simulation results and conclusions are presented in Section \ref{simu} and Section \ref{conc}, respectively.

{\em Notions:} All boldface letters indicate vectors (lower case) or matrices (upper case).  The superscripts   $(\cdot)^H$ and  $(\cdot)^{-1}$ denote the conjugate transpose and the inverse of a matrix, respectively. In addition, $\|\mathbf{z}\|_2$  denotes  the $L_2$  norm of a complex vector $\mathbf{z}$.  The operator $\mathcal{CN}(0, \mathbf{\Theta})$ represents a complex Gaussian vector with zero-mean and covariance matrix $\mathbf{\Theta}$. $\mathbf{I}_M$ denotes an identity matrix of size $M\times M$. Finally, $\leftarrow$ denotes the assignment operation.

\section{System Model, Problem Formulation and MAML Algorithm}\label{system_model}
\subsection{System Model and Problem Formulation}
We consider a multi-input   single-output (MISO) downlink  system where  a base station (BS) with $N_t$ antennas serves $K$ single-antenna users. The received signal at user $k$ can be written as
\begin{align}
\mathbf{y}_k = \mathbf{h}_k^H\mathbf{w}_k\mathbf{s}_k+  \mathbf{h}_k^H\sum_{j\neq k}^K \mathbf{w}_j\mathbf{s}_j +n_k,
\end{align}where $\mathbf{h}_k\in\mathbb{C}^{N_t\times1}$ denotes the channel vector between the BS and user $k$, $\mathbf{w}_k$ and $\mathbf{s}_k\sim\mathcal{CN}(0,1)$ denote the transmit beamforming vector and the information-bearing   signal for user $k$ with normalized power, respectively. The beamforming matrix $\qW=[\qw_1,\cdots, \qw_K]$ collects all beamforming vectors. The additive Gaussian white noise (AWGN) is given by $n_k\sim\mathcal{CN}(0,\sigma_k^2)$. As a result, the received SINR at user $k$ is expressed as
\begin{align}
\gamma_k = \frac{|\mathbf{h}_k^H\mathbf{w}_k|^2}{\sum_{j\neq k}^K|\mathbf{h}_k^H\mathbf{w}_j|^2+\sigma_k^2}, \forall k.
\end{align}
Based on the aforementioned system setup,  we consider a general utility maximization problem   under the total power constraint $P$  which is formulated as
\begin{align}\label{generalproblem}
\max_{\qW}~U(\gamma_1, \dots, \gamma_K), ~~~~\mathrm{s.t.}~~\sum_{k=1}^K\|\mathbf{w}_k\|_2^2\leq P,
\end{align}
where $U(\gamma_1, \dots, \gamma_K)$ is the utility, which is a non-decreasing function of individual users' SINR. We will consider maximization of the   SINR and the sum rate as two examples in Section IV.

Normally the problem \eqref{generalproblem} is nonconvex and can be solved by using iteration based-optimization algorithms such as \cite{schubert2004solution,bengtsson1999optimal} but with high computational complexity. Consequently, the computational latency of conventional optimization algorithms will  render  the beamforming solution obsolete when the CSI changes.
Although some existing deep learning-based algorithms \cite{shi2018learning,huang2018unsupervised,huang2019fast,xia2019deep} can reduce the complexity, they rely on the assumption that training data and testing data come from the same channel distribution to avoid serious performance deterioration caused by task mismatch issues. This strict assumption is difficult to satisfy in practical scenarios.  Fortunately,
because a common optimization problem with the same objective function and constraints will be solved in different channel conditions,   some inherent features exist in the solution structure of such an optimization problem which will help the beamforming design. Therefore, how to reuse such common features to design a fast meta learning based beamforming algorithm to cope with the dynamic wireless environments is the main objective of our paper.

\subsection{Overview of MAML}\label{theory}
 Although the MAML algorithm proposed in \cite{finn2017model} has been considered as an efficient optimization-based meta learning method for solving the mismatch issue, the complex optimization procedures also cause a high complexity. As a comparison to our proposed algorithm, we provide a brief overview  of the MAML algorithm.
\begin{figure}[htbp]
\centerline{\includegraphics[width=3in]{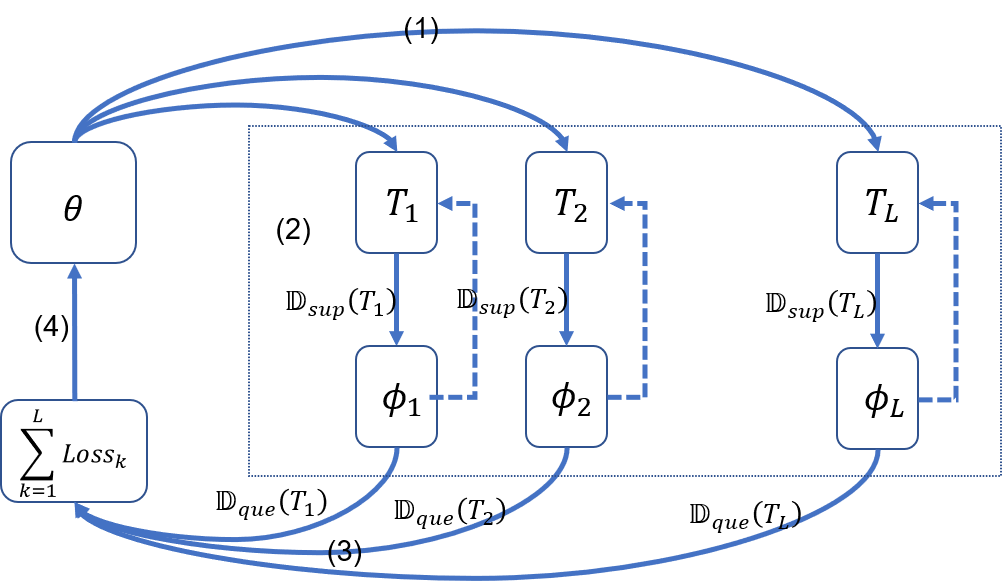}}
\caption{The workflow of the meta-training stage of the MAML algorithm: Step (1) -- network parameter initialization; Step (2) -- update of inner loop task parameters; Step (3) --  calculation of the loss function of  tasks on their query set; Step (4): update of the network parameter.}
\label{fig1}
\end{figure}

 The MAML algorithm aims to find a good neural network parameter initialization via two optimization loops in the meta training stage as shown in Fig. \ref{fig1}. The inner loop optimization procedure is used to optimize the task-specific parameters for each task based on the initialization of the neural network parameters as shown in step (2) of Fig. \ref{fig1}. The outer loop optimization procedure is used to optimize the initialization of the neural network parameters based on the task-specific parameters as shown in step (3) and (4) of Fig. \ref{fig1}. More specifically, we define the initial neural network parameters as $\mathbf{\theta}$, and the training task set as $\{\mathcal{T}(k)\}_{k=1}^{L}$, which includes $L$ tasks. For each task, the inner loop procedure aims to optimize its own task-specific parameters $\mathbf{\phi}_k$ on its support set $\mathbb{D}_{sup}(k)$, which can be expressed as
\begin{align}\label{taskoptipro}
\mathbf{\phi}_k=\arg\min_{\mathbf{\phi}_k}\mathrm{Loss}_{\mathbb{D}_{sup}(k)}(\mathbf{\phi}_k), \forall k,
\end{align}where $\mathrm{Loss}_{\mathbb{D}_{sup}(k)}$ is the loss function of task $k$. Based on a few steps of gradient descent, the task-specific parameters $\mathbf{\phi}_k$ can be updated by $\mathbf{\phi}_k^{(j)} = \mathbf{\phi}_k^{(j-1)}-\beta\nabla_{\mathbf{\phi}_k^{(j-1)}}\mathrm{Loss}_{\mathbb{D}_{sup}(k)}(\mathbf{\phi}_k^{(j-1)})$, where $j$, $\beta$, and $\nabla$  denote the iteration step, the learning rate, and the gradient function, respectively. Note that $\mathbf{\phi}_k^{(0)}$ equals to $\mathbf{\theta}$. Based on the task-specific parameters of each task, we define the meta-loss as $\mathrm{Loss}_{meta}(\mathbf{\theta})=\sum_{k=1}^L\mathrm{Loss}_{\mathbb{D}_{que}(k)}(\phi_k)$, where $\mathrm{Loss}_{\mathbb{D}_{que}(k)}(\phi_k)$ denotes the loss on the query set $\mathbb{D}_{que}(k)$ of task $k$ after the inner loop updates. Then, the initialization of the neural network parameters can be updated by minimizing the meta-loss via the outer loop optimization procedure, which is expressed as
\begin{align}\label{metaoptim}
\mathbf{\theta}=\arg\min_{\mathbf{\theta}}\mathrm{Loss}_{meta}(\mathbf{\theta}).
\end{align}By using the gradient descent update, the parameters $\mathbf{\theta}$ can be updated by $\mathbf{\theta}\!\leftarrow\!\mathbf{\theta}\!-\!\alpha\!\nabla_{\mathbf{\theta}}\mathrm{Loss}_{meta}(\mathbf{\theta})$, where $\alpha$ is the learning rate. Notice that the chain rule is included in calculating the parameters $\mathbf{\theta}$.

\section{Fast Meta Learning Framework and Applications}\label{fastalgorithm}
The complexity associated with the meta training and meta adaptation stages of the MAML algorithm is still high, which will affect how fast it can react to the changing environment. In this section, we aim to design a simple and efficient adaptation framework  that is able to achieve comparable performance as MAML. Our design is motivated by the observation in  \cite{tian2020rethinking} and \cite{raghu2019rapid} that a good   embedding model   that can extract key features is the most important factor to achieve effective adaptation.  In the following, we shall present the proposed fast meta learning framework  and its two applications to design adaptive downlink beamforming.

 \subsection{Design of Fast Meta Learning Framework}
\begin{figure}[h]
\centerline{\includegraphics[width=3.2in]{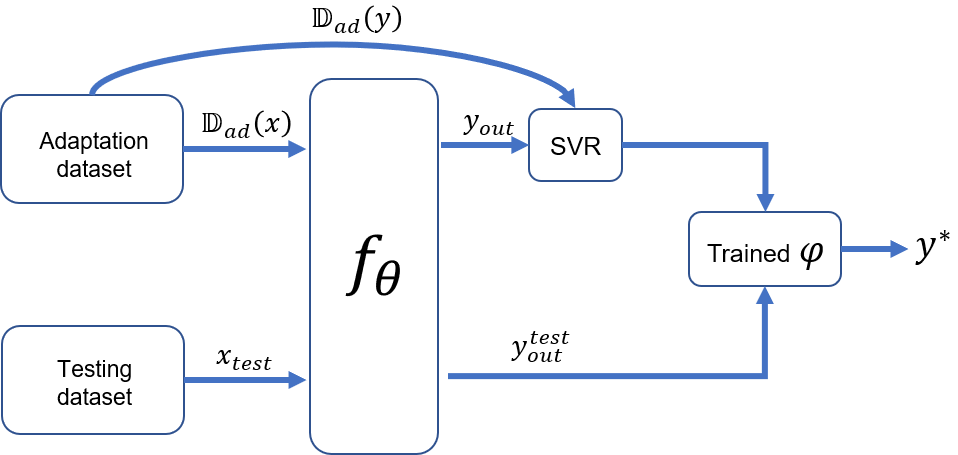}}
\caption{The workflow of the proposed fast adaptation algorithm.}
\label{fig2}
\end{figure}

The process of the proposed fast adaptation algorithmic framework is illustrated in Fig. \ref{fig2}, where $f_\theta$ is the pre-trained embedding model. Since feature reuse is the main reason for MAML to achieve good adaptation performance,  it is possible to train a feature extractor without multiple tasks and two optimization loops which are two factors that affect the training efficiency of MAML.

Rather than designing a new meta learning algorithm, we aim to design a simple way to train an embedding model for feature extraction based on a single task $\mathcal{T}$, which merges all meta training data $\mathcal{D}_{fast}=\cup\{\mathbb{D}_{sup}(k),\mathbb{D}_{que}(k)\}_{k=1}^{L}$. Based on the single task $\mathcal{T}$, the simple embedding model can be obtained by solving the following optimization problem
\begin{align}\label{fast1}
\theta=\arg\min_{\theta}\mathrm{Loss}_{\mathcal{D}_{fast}}(\theta),
\end{align}where $\mathrm{Loss}_{\mathcal{D}_{fast}}(\cdot)$ is the loss function defined as the difference between the predicted output and the target output. The parameters of the embedding model   can be updated using the gradient-based method as
\begin{align}\label{fast2}
\mathbf{\theta}\leftarrow\mathbf{\theta}-\alpha\nabla_{\mathbf{\theta}}\mathrm{Loss}_{\mathcal{D}_{fast}}(\theta).
\end{align}
Since the process of training the embedding model is similar to the conventional deep neural network training, which does not need to apply the alternating procedures, it is more time efficient compared to the model training of the MAML algorithm and this will be verified numerically. In addition,  depending on the specific problem, the training of the embedding model can use other methods such as Siamese networks and matching networks \cite{tian2020rethinking}, so our proposed embedding model based method is different from transfer learning and is    much more general.

As mentioned before, a direct employment of the pre-trained model to achieve the prediction on new tasks causes performance degradation,  and   fortunately feature reuse from the old tasks to the new tasks is able  to avoid such degradation. Hence, we can use the pre-trained embedding model to first extract the features of the new tasks. However, how to make use of the extracted features to achieve fast adaptation  is still a challenge. We propose to use SVR, which is a   fast regression algorithm based on the support vector machine, as a solution to post-process the extracted features for fast adaptation. The SVR technique aims to find a  hyperplane, which has the shortest distance to all data points. Specifically, the features of a new task are extracted by using the pre-trained embedding model over the associated adaptation dataset  $\mathbb{D}_{ad}$, which is expressed as
\begin{align}\label{fast3}
y_{out}=f_{\theta}(\mathbb{D}_{ad}(x)),
\end{align}where  $f_{\theta}$ denotes the pre-trained embedding model with the parameter $\theta$, $\mathbb{D}_{ad}(x)$ and $y_{out}$ denote the input dataset with $N_{ad}$ samples and the output features of the embedding model with dimensions of $N_{ad}\times 2\times N_tK$ and $N_{ad}\times K$, respectively.
$\mathbb{D}_{ad}(y)$ is the  final output data in the adaptation dataset with a dimension of $N_{ad}\times K$   associated with $\mathbb{D}_{ad}(x)$. Then $y_{out}$  and the associated labelled data ${\mathbb{D}_{ad}(y)}$ are used as input and output to train the SVR model to predict the final result.  The parameter $\varphi$ of the SVR model, which includes the weight $W$ and the bias $b$, can be obtained by minimizing the loss function below (which also includes a nonlinear kernel function), i.e.,
\begin{align}\label{fast4}
\varphi^*=\arg\min_{\varphi}\mathrm{Loss}_{\mathbb{D}_{ad}(y)}(W y_{out}+b,\mathbb{D}_{ad}(y)).
\end{align}

Once we have the pre-trained embedding model $f_\theta$ and   the SVR model denoted by $f_{\varphi^*}$, we can use them to find out the adaptive solution in the testing stage. Full details of the proposed fast adaptation solution are summarized in Algorithm 1.

 Compared to the MAML algorithm, our proposed fast meta-learning method has   two advantages.  First, the proposed fast adaptive algorithm has  lower complexity in both training and adaptation stages by using  a simpler training process and avoiding the fine-turning of the pre-trained model.  Let us discuss the complexity reduction in detail using the example of the neural network architecture in the next subsection. In the adaptation phase, for our proposed use of SVR model, there are $K^2+K$ parameters to optimize, while the MAML algorithm needs to optimize $FN^2(F+2)+2F +K^2N_tF+K$ variables which are the same as the original neural network, where $F$ is the size of the filter in the convolutional neural network (CNN) module.  Furthermore, the transfer learning method has $K^2 N_t F + K$ parameters to optimize corresponding to the last layer of the original neural network.  Second, our proposed method can be used in   either supervised or unsupervised learning in the pre-training stage as long as a good embedding model is obtained. In contrast, the MAML method relies on supervised learning and requires labelled data   in the pre-training stage.

\begin{table}[h]\label{Algorithm1}
\hrule
\vspace{1mm}
\noindent \textbf{Algorithm 1:}  The proposed fast adaptation framework for a regression problem. \label{Table: Table I}
\vspace{1mm}
\hrule
\vspace{1mm}
$\hspace*{2mm}$\textbf{Input:}  {Learning rate $\alpha$, meta training dataset $D_{fast}$, adaptation dataset $\mathbb{D}_{ad}$, testing dataset $\mathbb{D}_{test}$}\\
$\hspace*{2mm}$\textbf{Output:} Predict value $y^*$
\vspace{1mm}
\hrule
\vspace{2mm}
$\hspace*{4mm}$$\mathbf{Embedding~model~training}$
\vspace{-1mm}
\begin{enumerate}
\item Randomly initialize the network parameter $\mathbf{\theta}$
\item \textbf{while} not done \textbf{do}
\item $\hspace*{3mm}$ $\theta\leftarrow\theta-\alpha\nabla_{\theta}\mathrm{Loss}_{D_{fast}}(\theta)$ or by using the ADAM optimizer
\item \textbf{end while}
\end{enumerate}
\hrule
\vspace{2mm}
$\hspace*{4mm}$$\mathbf{Adaptation~and~testing}$
\vspace{-1mm}
\begin{enumerate}
\item $\mathbf{-----------  Adaptation-----------  }$
\item Extract the feature using the pre-trained embedding model $f_{\theta}$ on the adaptation dataset: $y_{out}=f_{\theta}(\mathbb{D}_{ad}(x))$
\item  Train the SVR model $\mathbf{\varphi}$ based on $y_{out}$ and $\mathbb{D}_{ad}(y)$  to obtain $f_{\mathbf{\varphi^*}}$
\item $\mathbf{------------ Testing------------ }$
\item Extract the feature using pre-trained embedding model $f_{\theta}$ on the testing dataset: $y_{out}^{test}=f_{\theta}(\mathbb{D}_{test}(x_{test}))$
\item Predict the output using the extracted feature: $y^*=f_{\mathbf{\varphi^*}}(y_{out}^{test})$.
\end{enumerate}
\hrule
\end{table}
\subsection{Applications}
In this subsection, the effectiveness of the proposed fast adaptive method will be verified by using two applications of utility maximization:
SINR balancing and sum rate maximization. The optimal solution of the former problem can be found by using numerical algorithms,  whereas it is not easy to find the optimal solution of the latter problem. Some iterative algorithms can be used to find   the sub-optimal solution of the sum rate maximization problem,  such as the weighted minimum mean squared error (WMMSE) algorithm \cite{shi2011iteratively,christensen2008weighted}. Therefore, their training methods for the embedding model are different, but we will show that both can be accommodated by the proposed framework.

\subsubsection{\textbf{SINR Balancing}}
Based on the  system model in Section II, the SINR balancing problem under the total power constraint $P$ can be formulated as \begin{align}\label{SINRproblem}
\max_{\qW}~\min_{1\leq k\leq K}~\gamma_k, ~~~~\mathrm{s.t.}~~\sum_{k=1}^K\|\mathbf{w}_k\|_2^2\leq P.
\end{align}

 Directly predicting beamforming causes high training complexity and inaccurate results due to the high dimensional beamforming matrix.  As mentioned before,  feature reuse is a key concept to achieve good adaptation performance. Therefore,  we propose to predict the low dimensional uplink power allocation vector as the main feature vector based on which the original beamforming matrix can be readily recovered.  According to the uplink-downlink duality in \cite{schubert2004solution} and \cite{xia2019deep}, using the uplink power allocation vector to recover beamforming   is possible because the same SINR region of the uplink and downlink problems can be achieved given the same total transmit power. Based on uplink-downlink duality and by defining the normalized beamforming $\mathbf{w}_k=\tilde{\mathbf{w}}_k\sqrt{p_k}$,     the downlink problem \eqref{SINRproblem} can be converted into the following uplink problem
\begin{align}\label{uplproblem}
\max_{\mathbf{q}}~\min_{1\leq k\leq K}~\frac{q_k|\mathbf{h}_k^H\tilde{\mathbf{w}}_k|^2}{\sum_{j\neq k}^Kq_j|\mathbf{h}_j^H\tilde{\mathbf{w}}_k|^2+\sigma_k^2},\notag \\
~~~~\mathrm{s.t.}~~\|\mathbf{q}\|_1\leq P,~\|\tilde{\mathbf{w}}_k\|_2=1,\forall k,
\end{align}where $\mathbf{q}=[q_1,\ldots,q_K]^T$ and $q_k$ is the uplink power allocation for user $k$, $\tilde{\mathbf{w}}_k$ and $p_k$ are the normalized beamforming and downlink power allocation of user $k$, respectively. With the predicted uplink power allocation vector, the original downlink beamforming can be recovered as follows.

First, the normalized beamforming vector can be obtained as $\tilde{\mathbf{w}}_k=\frac{(\sigma_k^2\mathbf{I}+\sum_{k=1}^Kq_k\mathbf{h}_k\mathbf{h}_k^H)^{-1}\mathbf{h}_k}
{\|(\sigma_k^2\mathbf{I}+\sum_{k=1}^Kq_k\mathbf{h}_k\mathbf{h}_k^H)^{-1}\mathbf{h}_k\|_2},\forall k$ and then the optimal downlink power allocation vector $\mathbf{p}=[p_1,\ldots,p_K]^T$ can be obtained by using the eigenvalue decomposition as detailed in the duality result in \cite{schubert2004solution}.  To be specific,
 the optimal downlink power allocation vector $\mathbf{p}$ is  obtained by finding the first $K$ components of the principal eigenvector of the following matrix
\begin{align}
\mathbf{\Upsilon}(\tilde{\mathbf{W}}, P)=\left[
\begin{matrix}
\mathbf{D}\mathbf{U}&\mathbf{D}\mathbf{\sigma}\\
\frac{1}{P}\mathbf{1}^T\mathbf{D}\mathbf{U}&\frac{1}{P}\mathbf{1}^T\mathbf{D}\mathbf{\sigma}
\end{matrix}\right],\nonumber
\end{align}where $\mathbf{1}=[1,1,\ldots,1]^T$, $\mathbf{D}=\mathrm{diag}\{1/|\tilde{\mathbf{w}}_1^H\mathbf{h}_1|^2,\ldots,1/|\tilde{\mathbf{w}}_K^H\mathbf{h}_K|^2\}$, $\mathbf{\sigma}=[\sigma_1^2,\sigma_2^2,\ldots,\sigma_K^2]^T$, and $[\mathbf{U}]_{kk^{'}}=|\tilde{\mathbf{w}}_{k^{'}}^H\mathbf{h}_k|^2$, if $k^{'}=k$, otherwise $[\mathbf{U}]_{kk^{'}}=0$.
Finally, the downlink beamforming matrix $\mathbf{W}=[\mathbf{w}_1,\ldots,\mathbf{w}_k]$ is derived as $\mathbf{W}=\tilde{\mathbf{W}}\sqrt{\mathbf{P}}$, where $\tilde{\mathbf{W}}=[\tilde{\mathbf{w}}_1,\ldots,\tilde{\mathbf{w}}_K]$ and $\mathbf{P}=\mathrm{diag}(\mathbf{p})$. The main advantages of predicting the uplink power allocation vector $\mathbf{q}$ rather than the original beamforming matrix  are to improve accuracy and to reduce complexity by reducing the output dimension from $2N_t K$ (the number of real-value variables in the beamforming matrix) to $K$.

Using the above analysis, we choose the uplink power as the output of the neural network. The sample pairs in the meta training dataset $D_{fast}$  include  the input channels and output uplink power allocation vectors, which are generated by solving the problem in \eqref{uplproblem}. The sample pairs in the  adaptation and training datasets can be generated in a similar way. In the pre-training stage of Algorithm 1, the embedding model will be learned by using supervised training over $D_{fast}$, while the adaptation and testing processes will be carried out by using the datasets accordingly based on the description in Algorithm 1.

\subsubsection{\textbf{Sum Rate Maximization}}
Based on the  system model in Section II, the sum rate maximization problem under the total power constraint $P$ can be formulated as
\begin{equation}\label{p3}
\setlength{\abovedisplayskip}{3pt}
\setlength{\belowdisplayskip}{3pt}
 \ \max_{\mathbf{W}}\ \  \sum^K_{k=1} \log_2(1+\gamma_k), \ \
\text{s.t.}\  \sum^K_{k=1}||\mathbf{w}_k||_2^2\leq P.
\end{equation}
Different from the SINR balancing problem in \eqref{SINRproblem}, \eqref{p3} is a nonconvex and well-known challenging problem and
no practical  algorithm is available to find the optimal solution in an efficient way, so it is difficult to generate enough labelled data.  However, we will still be able to exploit  key features of the optimal solution to the sum rate maximization   to facilitate the adaptive algorithm design.

 According to the results in \cite{bjornson2014optimal}, the optimal downlink beamforming vectors  for the sum rate maximization problem follows the parameterized structure below
\begin{equation}\label{solution struc of sumrate}
  \qw_k^{\ast}=\sqrt{p_k}\frac{(\qI_N+\sum^K_{k=1}{\frac{q_k}{\sigma^2}\qh_k\qh_k^H})^{-1}\qh_k}{||(\qI_N+\sum^K_{k=1}{\frac{q_k}{\sigma^2}\qh_k\qh_k^H})^{-1}\qh_k||_2}, \forall k,
\end{equation}
where $q_k$ is the virtual uplink power and $\sum_{k=1}^K q_k=\sum_{k=1}^Kp_k=P$.  The solution structure in \eqref{solution struc of sumrate} provides the required features $\qq$ and $\qp$  for the beamforming design in the problem  \eqref{p3}.  To simplify the design, it is shown in  \cite{dl_sumrate} that  \eqref{solution struc of sumrate} can be further simplified as
\begin{equation}\label{solution struc of sumrate2}
  \qw_k^{\ast}=\sqrt{q_k}\frac{(\qI_N+\sum^K_{k=1}{\frac{q_k}{\sigma^2}\qh_k\qh_k^H})^{-1}\qh_k}{||(\qI_N+\sum^K_{k=1}{\frac{q_k}{\sigma^2}\qh_k\qh_k^H})^{-1}\qh_k||_2}, \forall k,
\end{equation}
with negligible performance loss by using the same uplink and downlink power, i.e., $\qq=\qp$. As a result, we can still use the uplink power as the main features in the training process as the SINR balancing problem.

 Unfortunately, there does not exist a low-complexity algorithm in the literature that can find the optimal $p^{\ast}_k$ and $q^{\ast}_k$ in \eqref{solution struc of sumrate} or $q^{\ast}_k$ in \eqref{solution struc of sumrate2}. In order to obtain the embedding model in the pre-training stage of Algorithm 1, we train the neural network in an unsupervised learning way without relying on labelled data, whose loss function takes the objective function of sum rate directly as the  metric, i.e.,
\begin{equation}\label{loss2}
  \text{Loss}=-\frac{1}{2KL}\sum^L_{l=1}\sum^K_{k=1} \log_2\left(1+\gamma^{(l)}_k\right),
\end{equation}
where $L$ is the batch size.

 In the adaptation stage,  a small number of labelled data are still needed. The WMMSE algorithm is a good candidate to find the locally optimal solutions \cite{shi2011iteratively,christensen2008weighted},  so we will use the WMMSE algorithm to generate the required labelled data, i.e., the uplink power vector $\qq$ for the adaptation stage. Based on the labelled data in the adaptation dataset, the  SVR model will be trained by minimizing the following loss function, which uses the MSE metric, i.e.,
\begin{equation}\label{loss1}
  \text{Loss}=\frac{1}{2LK}\sum_{l=1}^L{\left(||\underline{\qq}^{(l)}-\hat{\qq}^{*(l)}||^2_2\right)},
\end{equation}
where $\underline{\qq}^{(l)}$ denotes the power vector obtained from the WMMSE algorithm, and $\hat{\qq}^{*(l)}$ is the predicted result in the adaptation stage.

\begin{figure}[t]
\centering
\includegraphics[scale=0.34]{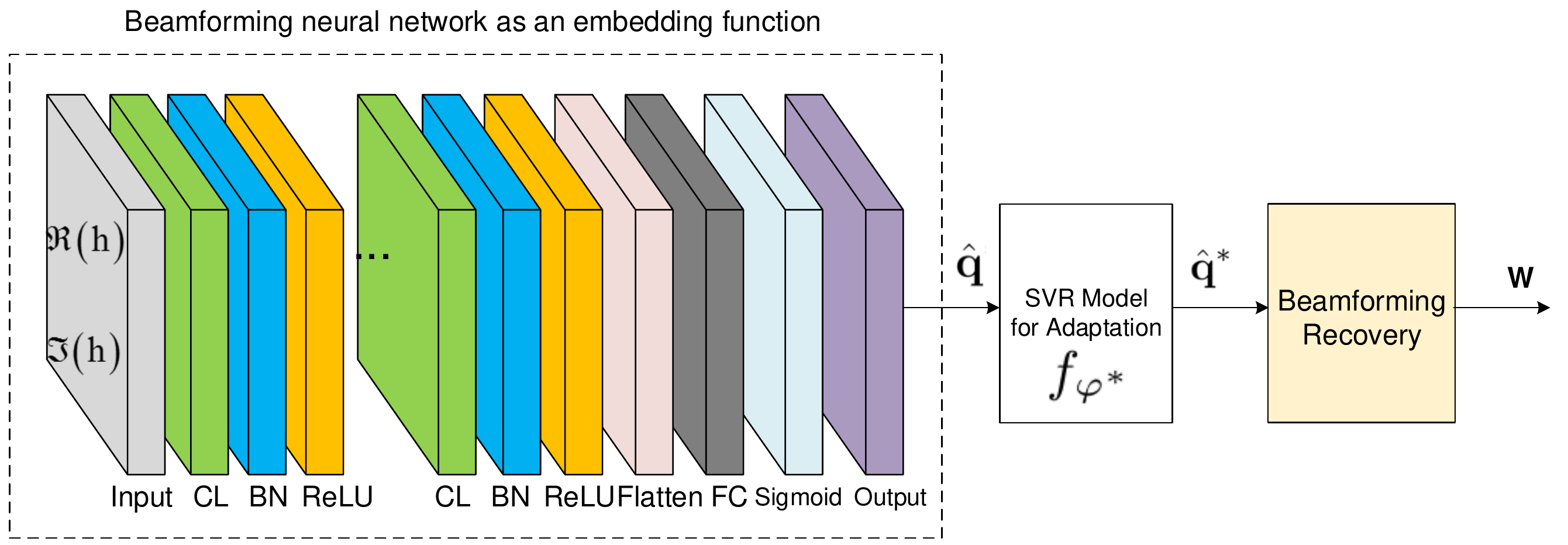}
\caption{An illustration of the proposed neural network architecture including the pre-trained embedding BNN model, the  SVR model for adaptation and beamforming recovery.}
\label{fig:NN}
\end{figure}
Before we conclude this section, we   provide details of the neural network architecture  to train the embedding model and adaptation model as illustrated in Fig. \ref{fig:NN}. Because both problems will extract the uplink power as the main feature, we can use a unified neural network architecture for embedding.
 We adopt the model-based beamforming neural network (BNN) framework proposed in \cite{xia2019deep}   as the network structure to obtain the embedding function $f(\theta)$. It takes the channel as input and the output feature is the uplink power vector $\hat \qq$.  This framework is composed of a  CNN architecture followed by the fully connected (FC) layer. CNN is chosen as the base of the embedding framework  due to its ability of extracting features and reducing learned parameters.  Specifically, the neural network used for our algorithm includes eleven layers: one input layer, two convolutional layer (CL) layers, two batch normalization (BN) layers, three activation (AC)  layers, one flatten layer, one FC layer and one output layer.    The complex channel input is split into two real value inputs, so the input dimension of the input layer is $2\times N_t K$. For the  two CL layers, each CL layer applies 8 kernels of size $3\times3$, one stride, and one padding. The input size of the first CL layer is equal to the size of the input data. The input size of the second CL layer and the output size of both CL layers are equal to $2\times N_t K\times 8$. Besides, ReLU and Sigmoid functions are adopted at the first two activation layers and the last activation layer, respectively. Adam optimizer is adopted \cite{kingma2014adam} for the optimization of the neural network. In the adaptation stage, the BNN output $\hat \qq$ is used as the input to train the SVR model, which then outputs the final adaptive uplink power vector $\hat \qq^*$. The beamforming recovery module is designed based on the uplink-downlink duality introduced in Section III.B-1.

\section{Online Application of Adaptive Algorithms}\label{online_adaptation}
Although the above proposed algorithm  can achieve the fast adaptive beamforming design based on the limited data of the new environment, they work in an offline manner assuming that the data used for adaptation is available in advance and the testing environment is stationary. They cannot cope with constantly changing situations such as vehicular communications and therefore online algorithms are needed.  This is because the channel data is likely to become available only sequentially, since effective channel estimation methods take time to first obtain the channel statistics and then estimate the channel; in addition, the channel distribution may be non-stationary as the environment keeps changing. Hence, it is necessary to design an efficient online adaptation algorithm to continuously adapt the model according to the changing environment in real-world applications. In the following, we will explain how to extend the idea of the proposed offline adaptation algorithm to the online scenario. As a comparison to our online adaptation algorithm, we will first provide a brief description about the online MAML learning method.
\subsection{MAML-based Online Learning}
 The purpose of online machine learning is to achieve continuous learning of prediction using sequential and non-stationary data, and the designed online algorithms should be able to update the best predictor for future data at each time slot. There exist some successful online learning algorithms, such as follow the leader (FTL) \cite{hannan1957approximation} and improved FTL \cite{shalev2012online}. Although FTL is a standard algorithm for online learning, it may not provide the efficient adaptation when the environment changes. The reason is that the principle of training a model using FTL at each time slot is similar to the joint training, which trains the model on a single task. In order to overcome the drawback of FTL when adapting to the new tasks, the authors in \cite{finn2019online} incorporate MAML into FTL to design the follow the meta leader (FTML) algorithm.

 Similar to the offline MAML algorithm introduced in Section \ref{theory}, the FTML algorithm still involves the inner and outer optimization loops in the training stage at each time slot. Since the channel data of each user is sequentially provided, it is impossible to divide tasks in advance like the offline manner. It means that the distribution of tasks is not known in advance. We define $\mathcal{T}_t$ as the task received at time slot $t$. Since the received task at the current time slot can only  be used for adaptation, there is no training process at the first time slot. As the data is accumulated in online learning, we define a task set $\mathcal{B}_t$ to store data of the task $t$ received at the time slot $t$, $t=0,\ldots, T$. After the task $\mathcal{T}_t $ is stored at time slot $t$, the algorithm will sample $N_{task}$ tasks from the previous task sets $\{\mathcal{B}_t,~t=0,\ldots, t-1\}$ as the training tasks. The $k$-th sampled task includes a support set and a query set. Based on the optimization process of inner loop introduced in Section \ref{theory}, the task-specific parameters of sampled tasks can be updated via a few step gradient descent. Then, the initialization of the neural network at time slot $t$ can be optimized by using the parameter updating procedure of the outer optimization loop.  Once the training process is complete, the task data of the current time slot stored in $\mathcal{B}_t, ~t=0,\ldots, T$ is used to adapt the trained model. The training and adaptation processes continue to provide online adaptation.

\subsection{The Proposed Fast Online Learning Algorithm}
Although the MAML-based online adaptation algorithm can efficiently overcome the mismatch issues in changing environment scenarios, the learning and adaptation processes are sophisticated and time-consuming due to the multiple tasks and two optimization loops. The complex processes may cause delay for real-time wireless communication applications. In order to achieve the rapid adaptation and to satisfy the latency requirements in the real-world communications scenarios, we extend the proposed fast adaptation method presented in Section III to the online scenario to design the fast online adaptive algorithm.

Since we have given the full   details of the proposed fast adaptive framework in Section \ref{fastalgorithm},   this subsection   will focus on how to extend it to   online applications and especially the adaptation stage.  The main idea of the proposed fast adaptation method is to  use the pre-trained embedding model to extract common features of the new tasks, and then   train a SVR model via the extracted features for adaptation. Before designing its online counterpart, we need firstly to define the datasets. In the online scenario, the data used for adaptation and testing arrives sequentially, which means that the  data used in the online learning is accumulated over time.  Hence, we define an empty buffer $\mathcal{B}_t$ to store the adaptation data and an empty set $D_t$ to receive the testing data at time slot $t$. In addition, we assume that the system is able to receive $N$ adaptation sample pairs and $U$ testing sample pairs of each user at each time slot. Note that the adaptation and testing data come  from the same distribution. According to different adaptation data used for feature extraction, we propose two online adaptation algorithms: 1) extracting features from adaptation data of the current time slot; 2) extracting features from all  adaptation data accumulated until the current time slot. Since the feature extraction and the training of SVR model for both algorithms are the same, we use the first algorithm, which uses the current time slot data for adaptation, to describe our proposed online adaptation algorithm. Specifically, we use   time slot $t$ as an example to describe the online adaptation process of the proposed algorithm. At the beginning of  time slot $t$, the algorithm uses buffer $\mathcal{B}_t$ to receive $N$ adaptation sample pairs for each user. Then, based on   \eqref{fast3}, the extracted feature $\mathbf{\widehat{q}}_t^*$ via the input data in $\mathcal{B}_t$ can be expressed as
\begin{align}
\mathbf{\widehat{q}}_t^*=f_{\theta}(\mathcal{B}_t(\mathbf{h}_t)),
\end{align}where $f_{\theta}$ is the pre-trained embedding model, $\mathcal{B}_t(\mathbf{h}_t)$ denotes the input channel   stored in the buffer $\mathcal{B}_t$. Next, we will train the SVR model by using the extracted feature $\widehat{\mathbf{q}_t^*}$ and the target output $\mathbf{q}_t$ in $\mathcal{B}_t$. Note that $\mathbf{q}_t$ is the uplink power allocation for both SINR and sum rate maximization problems. Similar to \eqref{fast4}, the parameter of the SVR model for our specific regression problems can be expressed as
\begin{align}
\mathbf{\phi}_t=\arg\min_{\mathbf{\phi}_t}\mathrm{Loss}(\mathbf{W}_t\widehat{\mathbf{q}_t^*}+\mathbf{b}_t,\mathcal{B}_t(\mathbf{q}_t)),
\end{align} where $\mathbf{\phi}_t$ is the SVR model parameter with the weights $\mathbf{W}_t$ and bias $\mathbf{b}_t$ at time slot $t$, respectively, and $\mathcal{B}_t(\mathbf{q}_t)$ is the target value stored in $\mathcal{B}_t$ associated with the input $\mathbf{h}_t$. Once the training of the SVR model is completed, we can predict the current testing output based on the trained SVR model and the pre-trained embedding model. Compared to the MAML-based algorithm, the execution time of the proposed fast online adaptive algorithm is significantly reduced because there is no iterative updating procedure in the pre-training stage and adaptation is also simpler, which will be verified by the results in Section V.

As before, the SINR and sum rate maximization problems are   considered as two applications to investigate the adaptation ability of the designed fast online algorithm. Both applications take the channel as the input and  the uplink power $\qq$ as the output. The difference is that the embedding model for the SINR problem is trained by   supervised learning while for the sum rate problem,  it is trained by   unsupervised learning.

\section{SIMULATION RESULTS}\label{simu}
In this section, numerical simulations are carried out to evaluate the advantages of the proposed adaptive beamforming   algorithms for different wireless communications scenarios.  We consider an MISO downlink with one BS and multiple users operating at the carrier frequency of 2.9 GHz, with 20 MHz bandwidth and noise power spectral density of -174 dBm/Hz. Other specific parameters including the number of antennas at the BS $N_t$, the number of users $K$ and the transmit power $P$ are provided in each figure. We employ the deep learning software  Keras  with Tensorflow  as the backend to pre-train the embedding model and Scikit-learn is used to train the SVR model.  We choose the nonlinear Gaussian radial basis function (RBF) kernel in the SVR training.  PyTorch  is adopted to implement the MAML related algorithms. All simulation results are generated by using a computer with an Intel i7-7700 CPU and a NVIDIA Titan Xp GPU.

In our simulation, the labelled data used for training the embedding model and the SVR model are generated by using three small-scale fading channel models in order to enrich the training dataset: Rayleigh model with distribution $\mathcal{CN}(\mathbf{0}, \mathbf{I}_M)$, Ricean model with the Ricean factor   3, and Nakagami model with a fading parameter   5 and an average power gain   2.  For each of the three fading models, we generate 5000 channel samples and obtain 5000 corresponding labelled data for  the SINR balancing problem using the algorithm in \cite{schubert2004solution}. Hence, the training dataset includes 15,000 sample pairs. For the sum rate maximization problem, we generate 10,000 channel samples for each fading model by using the WMMSE algorithm in \cite{shi2011iteratively}.

We consider the following  typical fading scenarios for testing the adaptation capability of the proposed learning algorithms.
\begin{itemize}
\item Large-scale fading case:   the path loss  is given by $\mathrm{PL}=128.1+37.6\log_{10}(d ~[km])$, where $d$ is the distance between a user and BS. The shadow fading follows the log-normal distribution with zero mean and $8$ dB standard deviation. The   small-scale fading follows the Rayleigh distribution with zero mean and unit variance.
\item WINNER II outdoor case: this is a typical fixed urban scenario specified in   WINNER II \cite{bultitude20074}.
\item Vehicular case: this is an urban vehicle-to-infrastructure (V2I) scenario defined in Annex A of 3GPP TR 36.885  \cite{3gpp885}.
\end{itemize}

For comparison, we introduce other three benchmarks, namely, the optimal solution, the MAML solution, and the non-adaptive learning solution. The definitions of all solutions for comparison are listed below.
\begin{itemize}
\item The optimal/suboptimal solution: for the SINR balancing problem,  the solution is  obtained by using the iterative algorithm proposed in \cite{schubert2004solution}; for the sum rate maximization problem,  we consider the WMMSE solution  in \cite{shi2011iteratively}, which is obtained by using an iterative algorithm. It serves as a performance upper bound for all other schemes.
\item The MAML solution: this solution shows the adaptation result of the existing MAML algorithm  \cite{finn2017model} described in Section II-B and Section IV-A for the offline and online scenarios, respectively.
\item  The transfer learning solution: this solution shows the adaptation result of the existing transfer learning algorithm  \cite{finn2017model} that fine tunes the last layer of the neural network to achieve the adaptation.
\item The non-adaptive solution: this solution shows the result of using a pre-trained model to predict the beamforming results directly in a different testing environment without any adaptation.
\end{itemize}

Below we present the results and analysis for the two applications.
\subsection{SINR Balancing}
\begin{figure}[t]
\centering
\includegraphics[width=3in]{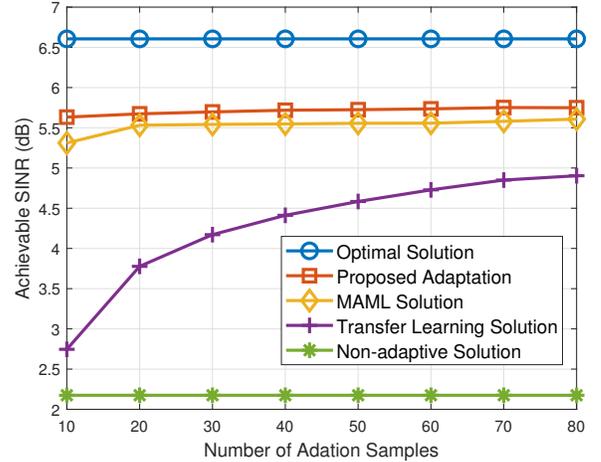}
\caption{The effect of the number of fine-tuning samples on the adaptation performance  when $N_t = 8, K = 8$, transmit power is 25 dBm.}
\label{fig:num:sample:SINR}
\end{figure}

We first investigate the large-scale fading case in which the BS is located at the center and all users are randomly distributed within a radius of $500$ m. First to decide how many samples will be used in the adaptation, we first study the effect of the number of fine-tuning samples for the same system. Note that more samples will certainly improve the adaptation performance, but also increase the computational complexity and latency. As shown in Fig. \ref{fig:num:sample:SINR},   SINR increases when the number of fine-tuning samples increases for both MAML and our proposed algorithms  and the SINR   converges fast for both algorithms. As a result, in the subsequent simulations, we consider 20 adaptation samples as a tradeoff between  a low adaptation overhead and satisfactory SINR performance.

\begin{figure}[t]
\centering
\includegraphics[width=3in]{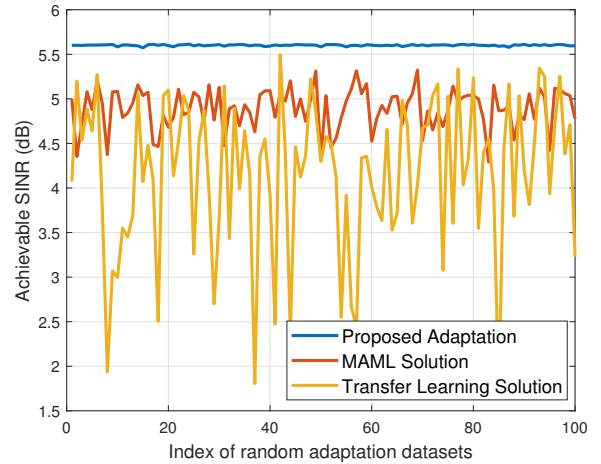}
\caption{Sensitivity of the SINR performance versus random adaptation datasets when $N_t = 8,K = 8,P = 25$ dBm.}
\label{fig:num:sensitivity:SINR}
\end{figure}

 In our simulation, we have observed that the performance of MAML and transfer learning solutions could vary depending on the adaptation datasets, so next we  study the sensitivity of different algorithms to the random adaptation datasets for an 8-user, 8-antenna system with 25 dBm transmit power in Fig. \ref{fig:num:sensitivity:SINR}, by using 100 random adaptation datasets each with a size of 20 samples. We can see that our proposed algorithm achieves robust SINR performance with different adaptation datasets but the achievable SINR of the transfer learning solution varies dramatically and the performance of MAML also fluctuates. In order to obtain stable performance comparison, we average the performance of MAML and transfer learning using 15 random adaptation datasets while for our proposed algorithm, we use just one random adaptation dataset.

\begin{figure}
\centering
\subfigure[]
{
	\begin{minipage}{3in}
    \label{fig2:a}
    \centering
	\includegraphics[width=3in]{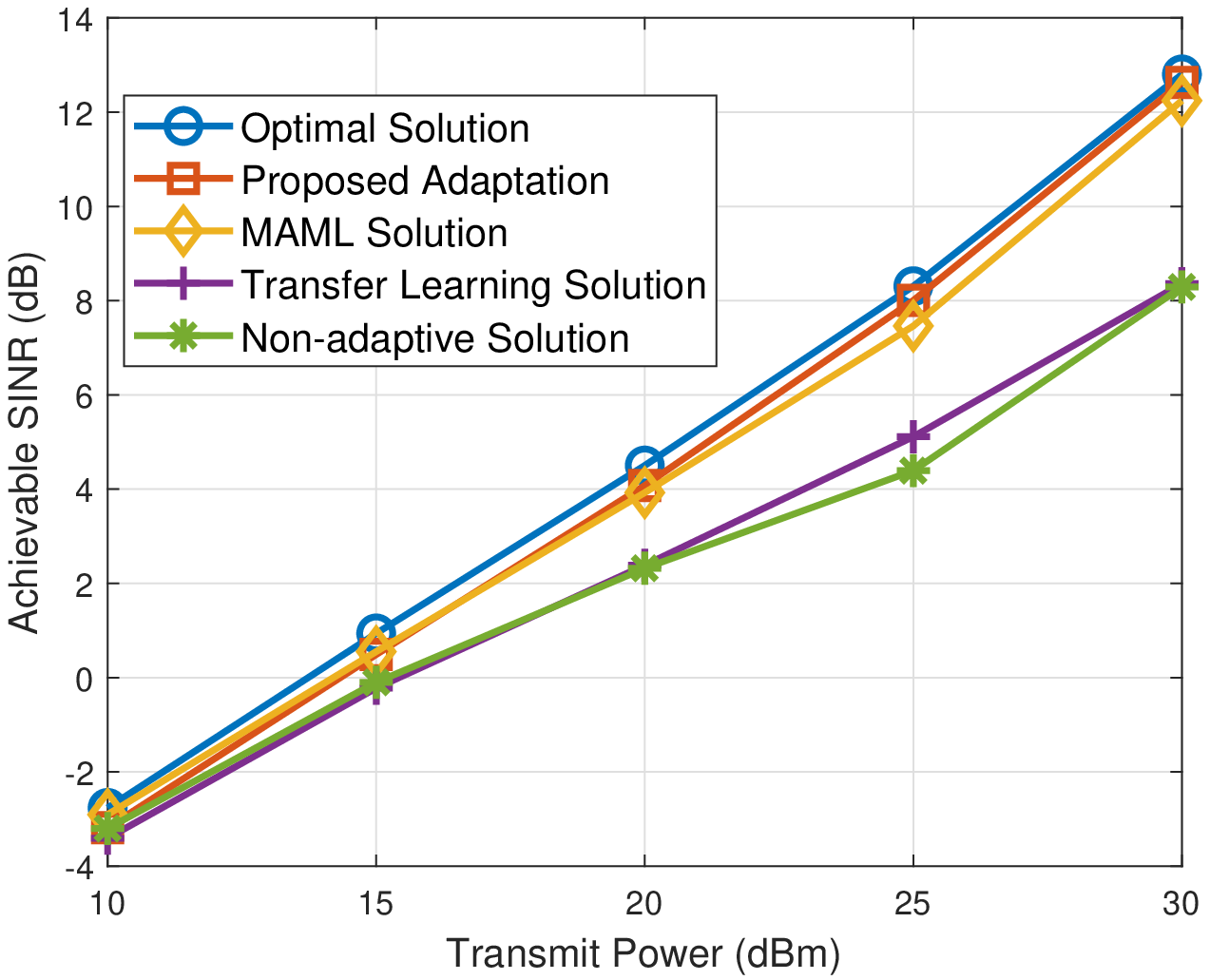}

	\end{minipage}
}
\subfigure[]
{
	\begin{minipage}{3in}
    \label{fig2:b}
    \centering
	\includegraphics[width=3in]{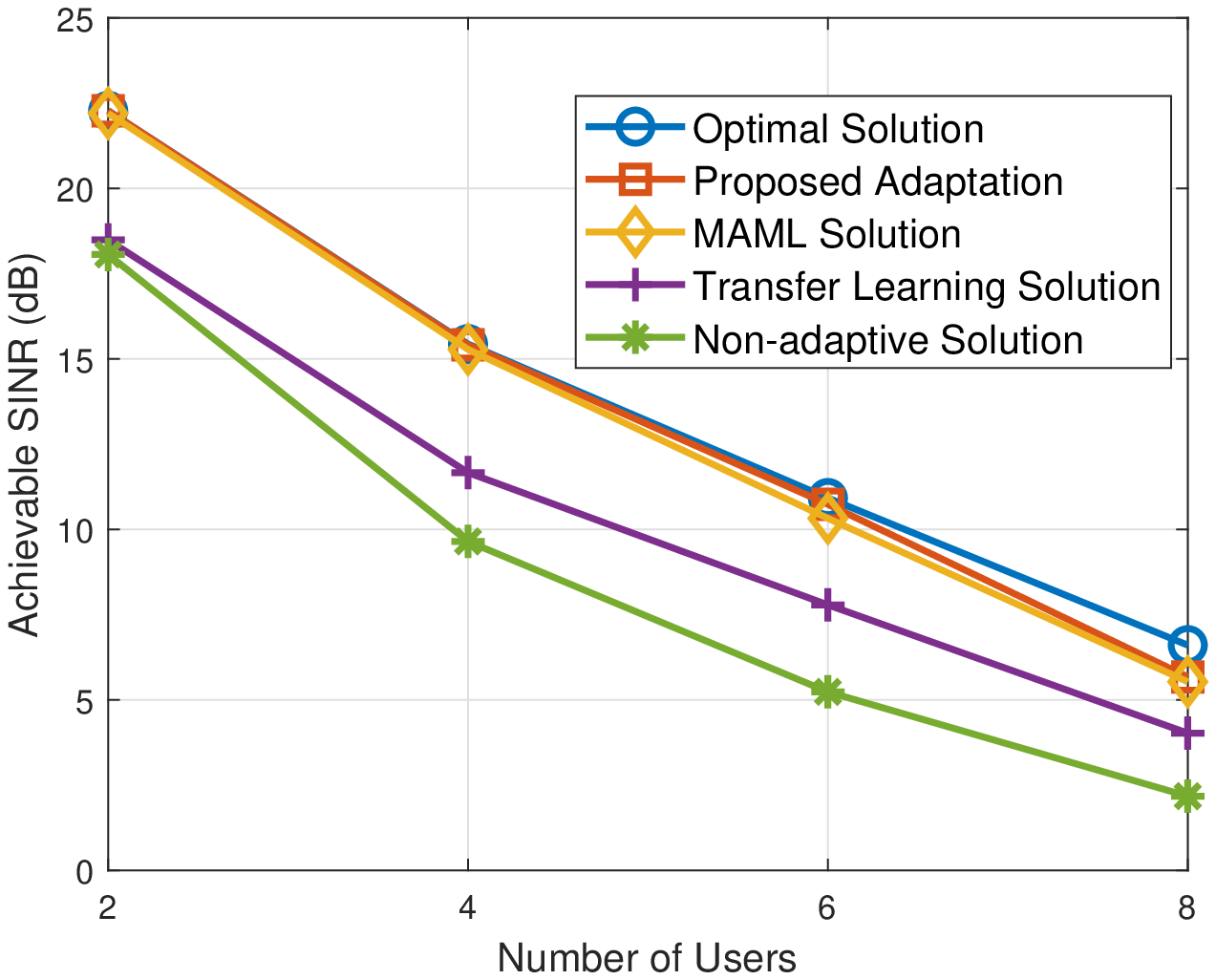}
	\end{minipage}
}
\centering
\caption{The SINR performance comparison in the large-scale  fading case versus: (a) transmit power when $N_t= 4$, $K = 4$ and (b) the number of users when $N_t= 8$, $P = 25$ dBm.}
\label{fig_large}
\end{figure}

\begin{figure}
\centering
 	\includegraphics[width=3in]{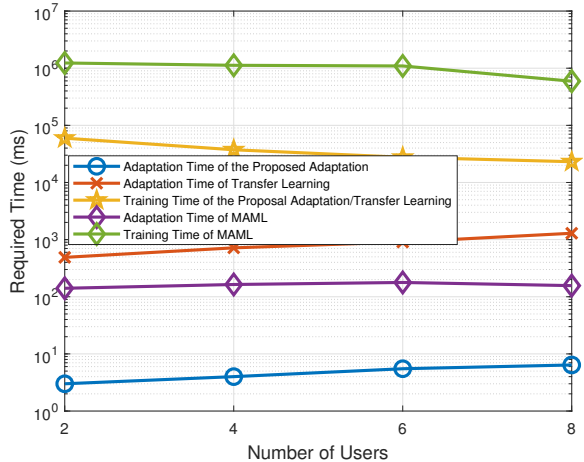}
 \caption{Comparison of the execution time of the adaptive algorithms.}
 \label{fig_time}
\end{figure}

  We then examine   the adaptation performance of the proposed algorithm   in Fig. \ref{fig_large}.  Fig. \ref{fig2:a} shows the effects  of the transmit power on the SINR performance. As expected, the achievable SINR improves as the transmit power increases for all schemes. The SINR performance associated with the proposed  algorithm is very close to that of the optimal solution and is better than the MAML solution. The non-adaptive solution achieves the worst SINR performance compared to the adaptive schemes.  The performance of the transfer learning solution is only slightly better than the non-adaptive solution.  In Fig. \ref{fig2:b}, the SINR performance versus the number of users is shown. It is observed that as the number of users increases, the performance gap between the MAML algorithm and the optimal solution is enlarged, but our proposed algorithm still achieves better performance. The results in  Fig. \ref{fig_large} verify that the proposed fast   algorithm provides an efficient adaptive beamforming solution.
  We compare the execution time of the adaptive algorithms in Fig. \ref{fig_time} versus the number of users. It can be seen that our proposed algorithm achieves more than an order of magnitude gain in terms of both training time and adaptation time compared to the MAML solution and therefore achieves faster adaptation.

\begin{figure}
\centering
 	\includegraphics[width=3in]{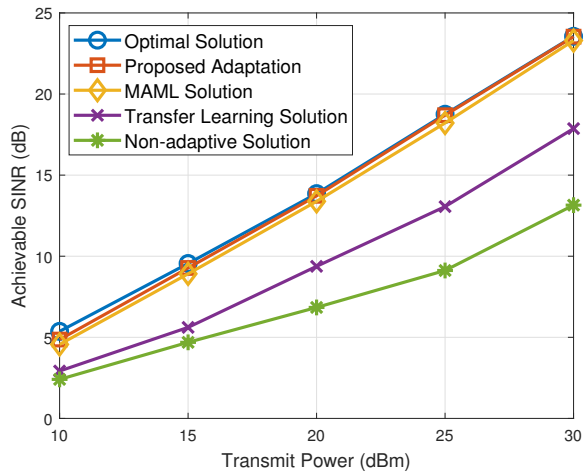}
 \caption{The SINR performance comparison in the WINNER II outdoor case when $M = 4$, $K = 4$.}
 \label{fig_outdoor}
\end{figure}

 To get a more accurate comparison of complexity between different algorithms, in Table \ref{table1} at the top of the next page, we show the numbers of passes of the neural network. For the pre-training and adaptation stages, each pass includes 100 and 20 channel data, respectively. We have adopted the early stop technique so the number of passes may not change monotonically as the number of users increases. It can be seen that the training of the proposed algorithm requires significantly less passes of the neural network than the MAML algorithm in the pre-training stage.
\begin{table*}[t]
\centering
\caption{Average number of passes of the neural network vs the number of users for the SINR balancing problem.}
\label{table system_configuration}
\begin{tabular}{|l|l|l|l|l|}
\hline
\textbf{Different Algorithms}  & \textbf{K=2} & \textbf{K=4} & \textbf{K=6} & \textbf{K=8}                                                                                                                \\ \hline
Pre-training of MAML   &                 130000      & 120000 &  115000       &60000
\\ \hline
Pre-training of adaptive algorithm/transfer learning     &   16950 & 10500 & 7350 & 6150                                                                                                                  \\ \hline
Adaptation of MAML     &      49                 &50 & 51&   73                                                                                                               \\ \hline
Adaptation of transfer learning     &   25                     & 47 & 60 & 87                                                                                                                   \\ \hline
\end{tabular}
\label{table1}
\end{table*}
Next we consider  the WINNER II outdoor scenario. We assume that the BS is located in the cell center and covers a disc with a radius of $1000$ m. Users are randomly distributed between $100$ m to $1000$ m away from the BS.   Fig. \ref{fig_outdoor} demonstrates the adaptation capability of the proposed learning algorithm  in the WINNER II outdoor case through the SINR performance. As can be seen from Fig. \ref{fig_outdoor}, the proposed   algorithm achieves slightly better  SINR performance than the MAML solution and both adaptive solutions significantly outperform  the non-adaptive solution.

\begin{figure}
\centering
 	\includegraphics[width=3in]{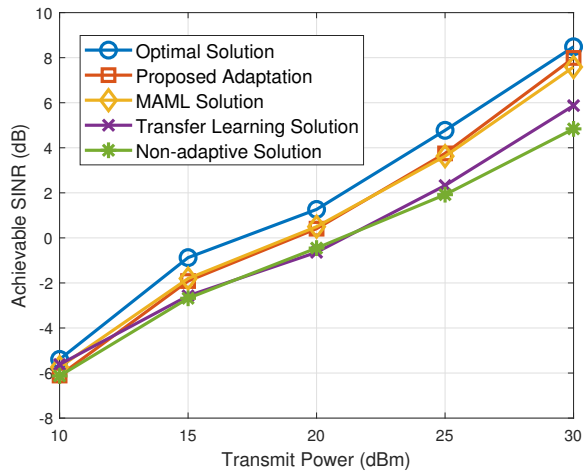}
 \caption{The SINR performance comparison in the urban case  when $N_t = 4$, $K = 4$.}
 \label{fig_urban}
\end{figure}

Next, we investigate the performance of the proposed algorithms for     V2I case in the urban environment, where we use the Manhattan grid layout with the region size of $750 ~m\times 1299 ~m$  and   grid size of  $250 ~m\times 433 ~m$. There are two 3.5m-wide lanes in each direction;  the BS is located in the center of the layout, and the vehicles are uniformly placed on each direction of the road. The probability of each vehicle to change its direction at the intersection is set to 0.4.  The velocity of each vehicle is 60 km/h. As seen in  Fig.  \ref{fig_urban},
the SINR performance generated by the non-adaptive solution is close to that of adaptive solutions, when the available transmit power budget is low; however, the advantages of the adaptive solution are gradually manifested as the available transmit power budget increases.
The possible reason for this interesting observation is that both adaptive and non-adaptive solutions require power to combat the negative effects of the fading channels and there is no enough power left for algorithms to efficiently achieve adaptation at the low power regime.  As the transmit power increases,   the proposed adaptive algorithm achieves   slightly better   performance than the MAML solution and much higher SINR performance than the non-adaptive solution.

Next, we evaluate the performance of the proposed online adaptive algorithm in real-world non-stationary scenarios. We investigate the adaptation capability of the proposed online fast adaptation algorithm in   changing environments by considering mobile users travelling from outdoor to urban and then highway environments.
The freeway case introduced in 3GPP TR 36.885 \cite{3gpp885} is used to generate channel data for the highway scenario. The number of lanes in each direction and the velocity of each vehicle are set as $3$ and $120$ km/h, respectively. The antenna gains of the BS and vehicles in the urban and highway scenario are set as 8 dBi and 3 dBi, respectively.
\begin{figure}[ht]
\centering
\includegraphics[width=3in]{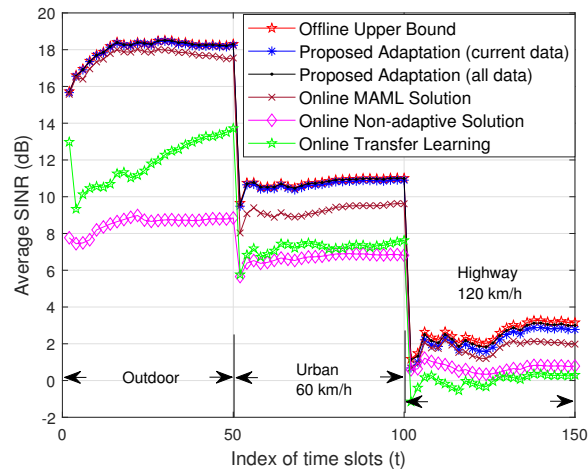}
\caption{Performance comparison between the proposed online algorithm and the existing algorithms {$M = 4$, $K = 4$, $P = 25$ dBm}.}
\label{SINR:online}
\end{figure}
Fig. \ref{SINR:online} shows the adaptation performance comparison between the proposed   fast adaptation algorithm and the MAML algorithm over the whole communications period as the users move across different environments. Each simulation point   is obtained by averaging all of the actual experimental points at the individual time slots over the previous time slots in the corresponding communication  scenario. To implement the simulation, we assume that five adaptation channels $N=5$ and ten testing channels $U=10$ of each user are received at each time slot. Each communications scenario lasts 50 time slots.  The offline fast upper bound solution shows the results obtained based on the fast algorithm in the offline manner where the embedding model and the  SVR model are trained by using the data of the corresponding communication  cases, respectively. The results of two online fast adaptations are obtained by using the data of the current time slot and all available data in the buffer, respectively. As we can see, there exists a significant reduction on the average SINR for all solutions, when the users travel from outdoor to urban and to highway environments due to the changes of   fading distribution. This fact indicates that the communication  quality in the real-world can be seriously affected by the continuously changing environment. In the figure, the average SINR performance associated with the proposed online fast adaptation algorithms is very close to that of the offline fast upper bound solution, which validates its adaptation ability.  In addition, the proposed online fast adaptation algorithms outperform the MAML algorithm and the non-adaptive solution for the whole communications period. It indicates that the proposed algorithm  can achieve fast adaptation by efficiently making use of sequential data when the environment changes. It is worth noting that the gap between the online fast algorithms and the MAML algorithm significantly increases when the communications scenario changes from outdoor to urban, whereas the gap reduces when the communications scenario changes from urban to highway. The reason is that there exists obviously difference of channel statistics features between the static scenario (outdoor) and mobility scenario (urban and highway). This fact further verifies that the proposed online fast adaptation algorithm has   better adaptation ability when the environment changes abruptly. The online fast adaptation  that utilizes all available historical data  performs better than that uses only the current data, and this is  because  more data can train a better SVR model in the adaptation  stage.   Note that for the urban case   and  the highway case in the online scenario, the channel coherent times are approximately 6.2 ms and 3.1 ms, respectively. From Fig. \ref{fig_time} and Fig. \ref{SINR:online}, we can see that our proposed algorithm can finish adaptation within 10 ms or only a few coherent time slots, which confirms that it can adapt to the fast changing wireless environment  in real-time.

\subsection{Sum Rate Maximization}
 In this subsection, we will investigate the performance of the adaptive algorithms to maximize the sum rate. As stated before, there is no practical algorithm that can achieve the optimal beamforming solution for this problem, so we use unsupervised learning in the pre-training stage and use the WMMSE solution to generate labelled data for   training of the SVR model in the adaptation stage.
 To use the MAML algorithm as a benchmark solution, we also use the WMMSE algorithm to generate training data  in its  pre-training stage. For the non-adaptive solution, we consider both unsupervised training and supervised training using the data generated by the WMMSE solution. In the literature, it has been reported that   unsupervised training achieves higher sum rate \cite{xia2019deep} which will  also be verified in our results below.

\begin{figure}[t]
\centering
\includegraphics[width=3in]{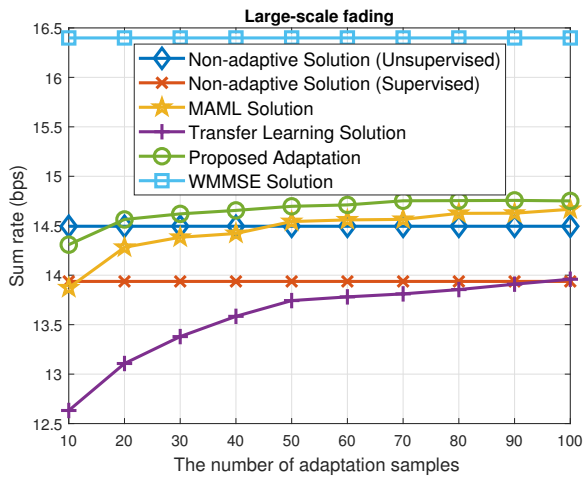}
\caption{The comparison of the number of adaptation samples when $M =  K = 4$, transmit power is 25 dBm.}
\label{sumrate:fig1}
\end{figure}

 We first examine the effects of the number of adaptation samples on the adaptive algorithms in Fig. \ref{sumrate:fig1} for a 4-user, 4-antenna system. Compared to the results of the SINR balancing problem in Fig. \ref{fig:num:sample:SINR}, we can see that   our proposed algorithm converges more slowly but it is still   faster than the MAML algorithm. As a tradeoff between the performance and the complexity, we choose 50 adaptation samples in the subsequent simulations.

\begin{figure}[t]
    \centering
	\includegraphics[width=3in]{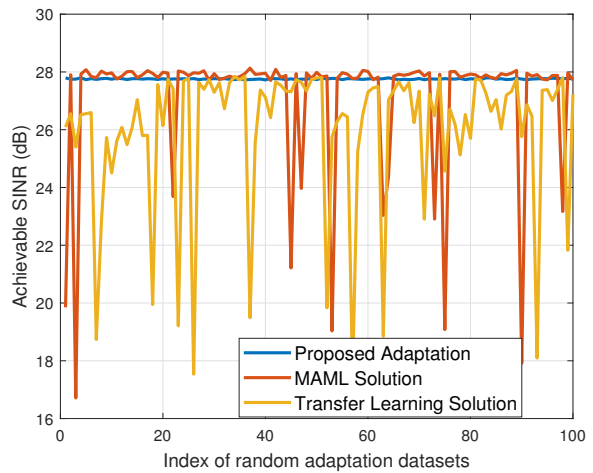}
\caption{Sensitivity of the sum rate performance versus random adaptation datasets when $N_t=8, K=8, P = 25$ dBm.}
\label{sumrate:random}
\end{figure}

 We next   investigate the sensitivity of the adaptive algorithms to the adaptation  datasets   in Fig. \ref{sumrate:random} for an 8-user, 8-antenna system with transmit power of 25 dBm. Fig. \ref{sumrate:random} shows that there exists a large variance on the sum rate results for both the MAML and the transfer learning solutions when adaptation dataset changes. It indicates that both of these two benchmark solutions heavily depend on the adaptation datasets. By contrast,  the proposed adaptive algorithm  demonstrates very robust performance when using different adaptation datasets. Similar to the SINR balancing problem, we average the results of the MAML and the transfer learning solutions using 15 random adaptation  datasets in order to obtain stable comparison results.

\begin{figure}
\centering
\subfigure[]
{
	\begin{minipage}{3in}
    \label{sumrate:fig2:a}
    \centering
	\includegraphics[width=3in]{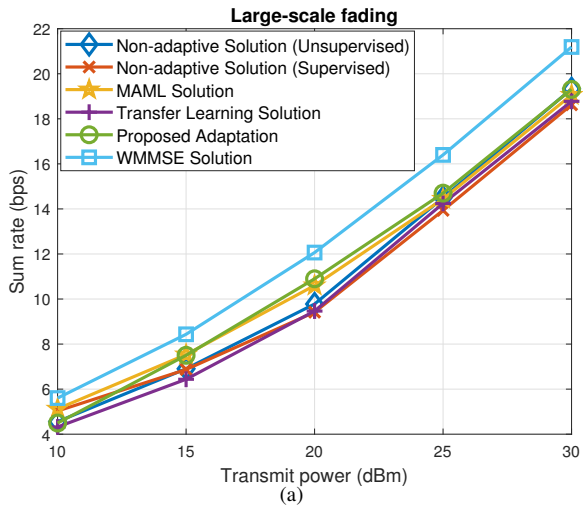}
	\end{minipage}
}
\subfigure[]
{
	\begin{minipage}{3in}
    \label{sumrate:fig2:b}
    \centering
	\includegraphics[width=3in]{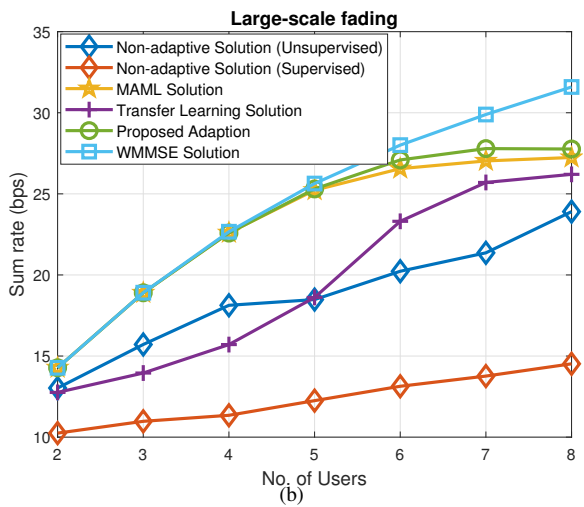}
	\end{minipage}
}
\centering
\caption{The sum rate performance comparison on large-scale case for different metrics: (a) transmit power when $N_t = 4$, $K = 4$ and (b) the number of users when $N_t = 8$, $P = 25$ dBm.}
\label{sumrate:large}
\end{figure}

The sum rate results in the large scale fading case are presented in Fig. \ref{sumrate:large}. As can be seen in Fig. \ref{sumrate:fig2:a}, for a 4-user 4-antenna system, our proposed solution achieves slightly better performance than the MAML solution, and the performance gap between different algorithms is not significant. However, Fig. \ref{sumrate:fig2:a} shows that as the number of users changes, the performance of transfer learning varies greatly.

\begin{figure}
\centering
 	\includegraphics[width=3in]{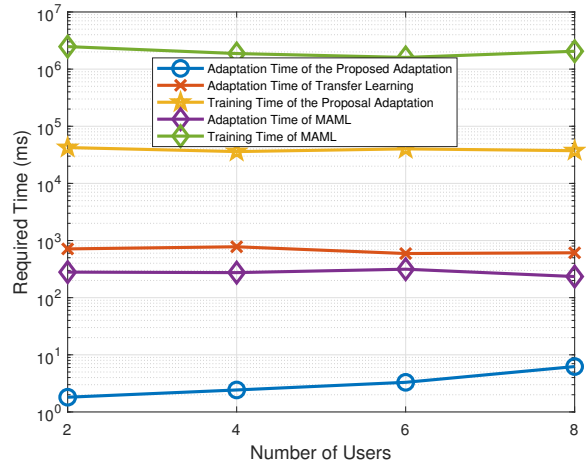}
 \caption{Comparison of execution time of adaptive algorithms for the sum rate maximization problem when $N_t=8, P = 25$ dBm.}
 \label{sumrate:fig_time}
\end{figure}

The execution time of the   adaptive algorithms is compared in Fig. \ref{sumrate:fig_time} for an 8-antenna system. It can be seen that the proposed algorithm uses less pre-training time than the MAML algorithm. It  finishes the adaptation in a few milliseconds and is three orders of magnitude faster than the MAML algorithm.
 Similar to Table \ref{table1},  we show the numbers of passes of the neural network in Table \ref{table2} at the top of the next page. It is observed again that the training of the proposed algorithm requires significantly less passes of the neural network than the MAML algorithm. The transfer learning solution also requires less passes than the MAML algorithm in the adaptation stage.
\begin{table*}[t]
\centering
\caption{Average number of passes of the neural network vs the number of users for the sum rate problem.}
\label{table system_configuration}
\begin{tabular}{|l|l|l|l|l|}
\hline
\textbf{Different Algorithms}  & \textbf{K=2} & \textbf{K=4} & \textbf{K=6} & \textbf{K=8}                                                                                                                \\ \hline
Training of MAML   &       255000             & 185000& 155000&200000
\\ \hline
Training of adaptive algorithm/transfer learning     &   7860 & 6600 & 7230 & 6570                                                                                                               \\ \hline
Adaptation of MAML     &          96              &102 & 98&  74                                                                                                                \\ \hline
Adaptation of transfer learning     &   52                      & 50 & 27 & 21                                                                                                              \\ \hline
\end{tabular}
\label{table2}
\end{table*}

\begin{figure}
\centering
 	\includegraphics[width=3in]{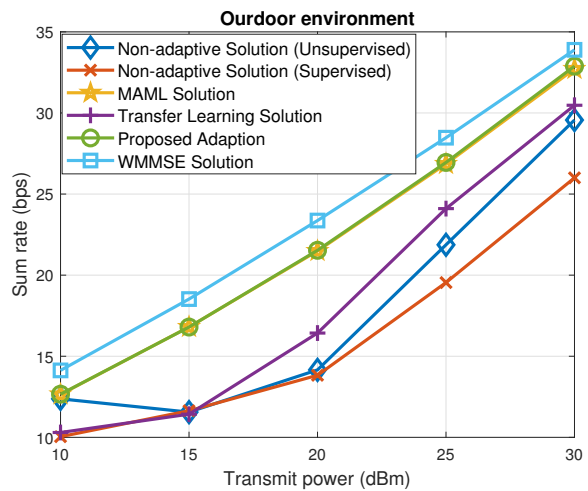}
 \caption{The sum rate performance comparison in the WINNER II outdoor case when $N_t = 4$, $K = 4$.}
 \label{sumrate:fig_outdoor}
\end{figure}

\begin{figure}
\centering
 	\includegraphics[width=3in]{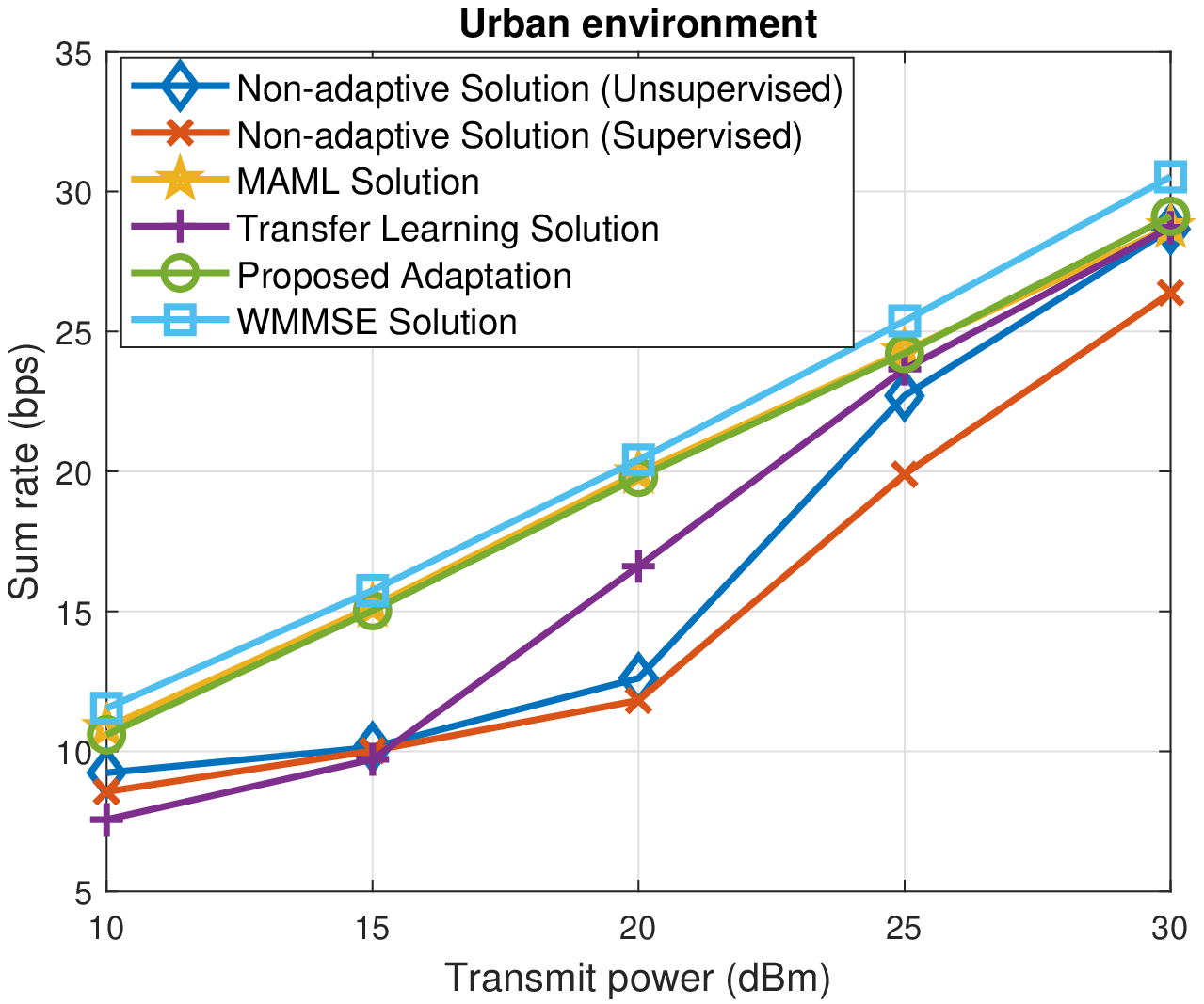}
 \caption{The sum rate performance comparison in the urban case  when $N_t = 4$, $K = 4$.}
 \label{sumrate:fig_urban}
\end{figure}

The  sum rate in the WINNER II outdoor scenario is compared in a 4-user, 4-antenna system. As shown in Fig. \ref{sumrate:fig_outdoor},
  both  adaptive algorithms achieve   good sum rate performance which are close to the results generated by the WMMSE solution.   Although the unsupervised learning based non-adaptive solution performs well in the low transmit power regime, its performance degrades fast as the transmit power increases. Similar performance comparison can be observed from the results in Fig. \ref{sumrate:fig_urban} for the V2I case in the urban environment, in which both adaptive solutions achieve the sum rate closer to that of the WMMSE solution than in the outdoor scenario.

\begin{figure}[ht]
\centering
\includegraphics[width=3in]{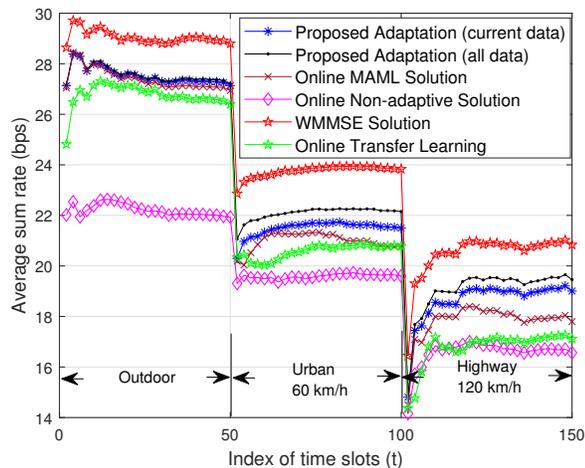}
\caption{Sum rate performance comparison between the proposed online algorithm and the existing algorithms, {$N_t = 4$, $K = 4$, $P = 25$ dBm}.}
\label{sumrate:online}
\end{figure}

Finally we evaluate the  sum rate performance of the  online algorithms in real-world non-stationary scenarios. Similar to the SINR maximization problem, we consider mobile users travelling from outdoor to urban and then highway environments.  Fig. \ref{sumrate:online} shows the adaptation performance comparison between the proposed   fast adaptation algorithm and the MAML algorithm over the whole communications period as the users move across different environments. To implement the simulation, we use the same number of adaptation channels and testing channels as in the SINR balancing problem. As before, the users' sum rate drops significantly when users cross borders of different environments.
  In the figure, the sum rate results achieved by the proposed online fast adaptation algorithm  are similar to that of the WMMSE solution and  outperform the MAML algorithm and the non-adaptive algorithm, which validates its adaptation ability in the changing environments. We note that the gap between the online fast algorithms and the MAML algorithm  increases when the communications scenario changes from outdoor to urban, and from urban to highway environments. This observation further confirms that   the proposed online fast adaptation algorithm has   better and faster adaptation ability when the environment changes.

\section{Conclusions}\label{conc}
In this paper, we proposed a simple and efficient adaptive   method for   beamforming design in dynamic wireless environments. The core idea of this method is to train a good model embedding  that extracts key features, followed by fitting a simple SVR model for fast adaptation. We have extended the proposed algorithm to online adaptation that performs well in constantly changing environments.  Simulation results demonstrated  that the proposed adaptive algorithms achieve a  better adaptation performance with a reduced  complexity compared to the existing meta learning algorithm. Our results   shed  some light on designing adaptive learning algorithms from few data for general resource management problems in wireless networks.  As to future work, we envisage that it would be an interesting direction to extend the proposed method to adapt to the varying number of active users, and the massive MIMO scenario in which  channel estimation and adaptive beamforming will be jointly learned using the deep learning approach.

 \end{document}